\documentclass{article}

\usepackage{arxiv}

\usepackage[utf8]{inputenc} 
\usepackage[T1]{fontenc}    
\usepackage{hyperref}       
\usepackage{url}            
\usepackage{booktabs}       
\usepackage{amsfonts}       
\usepackage{nicefrac}       
\usepackage{microtype}      
\usepackage{lipsum}
\usepackage{graphicx}
\graphicspath{ {./images/} }
\usepackage{array,multirow}
\usepackage{authblk}
\usepackage{amsmath}
\usepackage{amssymb}
\usepackage{mathtools}
\usepackage{dirtytalk}

\newtheorem{theorem}{Theorem}[section]

\newtheorem{definition}[theorem]{Definition}

\title{SoK: Differential Privacy on Graph-Structured Data}

\author[1,2]{Tamara T. Mueller}
\author[1,2]{Dmitrii Usynin}
\author[3,4]{Johannes C. Paetzold}
\author[1,2,5]{Daniel Rueckert}
\author[1,2,5]{Georgios Kaissis}

\affil[1]{Chair for AI in Medicine and Healthcare, Department of Informatics, Technical University of Munich}
\affil[2]{Department of Diagnostic and Interventional Radiology, Faculty of Medicine, Technical University of Munich}
\affil[3]{Department of Informatics, Technical University of Munich}
\affil[4]{Institute for Tissue Engineering and Regenerative Medicine, Helmholtz Zentrum München}
\affil[5]{Department of Computing, Imperial College London}

\begin{document}
\maketitle
\begin{abstract}
    In this work, we study the applications of differential privacy (DP) in the context of graph-structured data. We discuss the formulations of DP applicable to the publication of graphs and their associated statistics as well as machine learning on graph-based data, including graph neural networks (GNNs). The formulation of DP in the context of graph-structured data is difficult, as individual data points are interconnected (often non-linearly or sparsely). This connectivity complicates the computation of individual privacy loss in differentially private learning. The problem is exacerbated by an absence of a single, well-established formulation of DP in graph settings. This issue extends to the domain of GNNs, rendering private machine learning on graph-structured data a challenging task. A lack of prior systematisation work motivated us to study graph-based learning from a privacy perspective. In this work, we systematise different formulations of DP on graphs, discuss challenges and promising applications, including the GNN domain. We compare and separate works into graph analysis tasks and graph learning tasks with GNNs. Finally, we conclude our work with a discussion of open questions and potential directions for further research in this area. 
\end{abstract}


\section{Introduction}
\label{sec:introduction}
The utilisation of non-Euclidean data allows various learning contexts to benefit from additional structural information embedded in datasets such as graphs \cite{fan2019graph}, meshes \cite{luz2020learning}, and point clouds \cite{landrieu2018large}. In this work we focus on graphs in particular. Due to their inter-connected nature, relational information can be incorporated into the learning process, allowing the models trained on graph data to be deployed in contexts which rely on interactions between individual data points. Examples of such contexts include market value prediction \cite{matsunaga2019exploring}, fake news detection in social networks \cite{benamira2019semi} and drug development \cite{gaudelet2021utilizing}. Graph neural networks (GNNs) \cite{scarselli2008graph} have recently been proposed as an effective framework to perform machine learning over graph-structured data and have already been applied to various learning contexts \cite{igamberdiev2021privacy, olatunji2021releasing, sajadmanesh2020locally}.

However, while the aforementioned additional structural information allows the models trained on this data to incorporate the interactions between individual data points, such relational information (in addition to the data contained in the nodes itself) is often sensitive in nature. This renders analysing and/or conducting machine learning tasks on graphs problematic, as availability of such datasets cannot always be guaranteed due to various data protection regulations (e.g. the European General Data Protection Regulation \cite{radley-gardnerfundamental2016}).

Moreover, the emphasis on inter-node relationships, makes graph-structured data more vulnerable to attacks that target the privacy of individuals \cite{zhang2021graphmi, liu2016dependence}, intended to infer information that the individuals did not consent to disclosing. Such attacks can take form of membership inference (MIA) \cite{shokri2017membership}, where the adversary attempts to verify if a record that they possess was part of the sensitive dataset (e.g. a patient's electronic health record). MIA, in fact, has a higher fidelity in a graph-based settings, due to additional information that intrinsically lies in the structure of a graph \cite{olatunji2021membership}. Another commonly used attack is termed attribute (or feature) inference attack \cite{he2021node}, that aims to reconstruct sensitive features of individuals in the training dataset. This attack typically involves an adversary having access to a non-overlapping dataset of publicly available attributes which, alongside with the predictions of the trained model, are used to determine the value of a sensitive feature that belongs to a target participant. Furthermore, models trained on graph-structured data, such as GNNs, were shown to be susceptible to model inversion attacks (MInv) \cite{fredrikson2015model}, which allow the adversary to extract sensitive training data by leveraging the internal representations of the model (e.g. reverse-engineering a model update into disclosing which datapoint corresponds to this specific update). Authors in \cite{zhang2021graphmi} show that MInv attacks can be adapted to graph-based learning. Seeing as graph-structured data captures information not just about individuals themselves, but about their relationships with other participants, all of these attacks can potentially compromise privacy of multiple participants at once. 

Differential privacy (DP) \cite{dwork2014algorithmic} was proposed to objectively quantify the privacy loss of individuals whose data is subjected to algorithmic processing. Differentially private algorithms upper-bound the amount of information that can be inferred by the adversary from observing the computation's output, thus mitigating the attacks discussed above. The utilisation of DP mechanisms allows to train models on sensitive datasets while preserving privacy of contributors' data. 

However, a change of context to non-Euclidean spaces can affect the amount of information which can be extracted by adversaries and hence the formulation of what precisely is defined (and thus quantified) as private data. This implies that formal privacy enhancing technologies (PETs) need to be adapted to graphs in order to provide meaningful privacy protection. When considering DP on graph-structured data, such adaptation is non-trivial, as the decision about which formulation of DP that should be applied in each separate setting needs to consider not only the data of the learning task, but also the structure of the database. For instance, data owners need to define prior to commencing the learning task exactly which constituents of the graph are considered to be private. There exist formulations of DP that consider private information of individuals as being represented by the relationships between them (\textit{edge-level} DP). Other notions protect the individual features in nodes of the graph as well as their adjacent edges (\textit{node-level} DP) or whole graphs as one data entity (\textit{graph-level} DP). Some further definitions of DP on graph-structured data were derived from the ones above and discussed in more detail in Section \ref{sec:dp_graphs}.

As graph-structured data is more vulnerable to privacy-oriented adversaries and the application of DP is non-trivial in graph learning contexts, we identify a requirement for a comprehensive systematisation of existing knowledge in the area. In this work, we present an in-depth overview of differentially private graph analysis and learning on graph-structured data, outlining existing implementations, their limitations and application areas. We additionally outline a number of challenges associated with private learning on graph-based structures and promising directions for future work. Our contributions can be summarised as follows:

\begin{itemize}
    \item We systematise and discuss the existing formulations of DP in the context of graph analysis and graph learning, showing how different DP formulations can be applied in different contexts;
    \item We identify the limitations of these approaches and pinpoint promising areas of future work as well as open challenges in the domain of differential privacy on graph-structured data.
\end{itemize}

\section{Background}
\label{sec:background}
In this section, we give a brief introduction to graph neural networks (GNNs), formalise the concept of DP, introduce the three main notions on DP on graph-structured data, as well as the concept of sensitivity, the Gaussian and the Laplace mechanisms.

\subsection{Graph Neural Networks}
In multiple real-world scenarios, learning problems rely on non-Euclidean structures such as graphs, point clouds, or manifolds, rendering conventional deep learning models and approaches unsuitable. To mitigate this issue, GNNs were proposed \cite{scarselli2008graph} to leverage the full underlying structure of the dataset and maximize learning capacity by directly learning \textit{on the graph}. In the following, we will refer to a graph $\mathit{G} = \{\mathit{V}, \mathit{E}\}$ as the set of nodes $\mathit{V} = \{v_1, v_2, ..., v_n\}$ and edges $E = \{e_1, e_2, ..., e_m\}$, $n$ and $m \in \mathbb{N}$. Here, $n$ determines the number of nodes in the graph and $m$ the number of edges. We can classify graph learning scenarios into \textit{transductive} and \textit{inductive} applications. In an inductive learning setting the goal is to generalise from a specific training environment to universal rules, applicable in different testing environments. Transductive learning on the other hand, learns rules from specific observed training data to specific unobserved (test) data. The setting of transductive learning is often used in graph learning and referred to as semi-supervised learning. GNNs can be applied to either a single graph or a multi-graph dataset, depending on task and application. The three major application areas of GNNs are\textit{ node classification} (where one label is predicted for each node in the graph), \textit{edge prediction} (where edges are predicted or labeled), and \textit{graph classification} (where one label is predicted for each graph).

A key concept for the successful application of GNNs is message passing \cite{kipf2016semi}, where information is shared along edges and therefore propagated among neighbourhoods of nodes. This property enables the utilisation of the full dimensionality of graph datasets. However, this typically complicates the disentanglement of contributions by individual nodes, making the calculation of individual privacy loss per each participant a challenging task.

\subsection{Differential Privacy}
Differential privacy (DP) is a stability condition on randomised algorithms that makes it approximately invariant to an inclusion or exclusion of a single individual \cite{dwork2014algorithmic}. In the words of the authors of \cite{dwork2014algorithmic}, DP promises \say{to protect individuals from any additional harm that they might face due to their data being in the private database that they would not have faced had their data not been part of [the database]}. The DP framework and its associated techniques allow data analysts to draw conclusions about datasets while preserving the privacy of individuals. 

In a setting of DP on graph-structured data, we assume that an analyst $\mathcal{A}$ is entrusted with a database $D$ containing sensitive graph-structured data. From $D$ a neighbouring (in this work we additionally use the term \textit{adjacent}) dataset $D'$ is constructed by either (a) removing or adding one node and its adjacent edges (\textit{node-level} DP), (b) removing or adding one edge (\textit{edge-level} DP), or (c) removing or adding one graph (\textit{graph-level} DP).
Formally, DP can be defined as follows:

\begin{definition}[($\varepsilon$-$\delta$)-DP] \textit{A randomised algorithm $\mathcal{M}$ is ($\varepsilon$-$\delta$)-differentially private if for all $S \subseteq$ Range($\mathcal{M}$) and all neighbouring datasets $D$ and $D'$ in $X$ the following (symmetric) statement holds:}

\begin{equation} \label{eq:ed_dif_priv}
    \mathbb{P}[M(D) \in S] \leq e^{\varepsilon} \mathbb{P}[M(D') \in S] + \delta.
\end{equation}
\end{definition}

The definition of neighbouring datasets on graph-structured data depends on the desired formulation of privacy in the setting (i.e. which attributes need to be kept private, such as outgoing edges for instance).
Therefore, the desired notion (as well as the associated mechanisms) of privacy preservation depend on what the data owner requires to protect, the structure of the graph and the desired application. In order to employ differentially private algorithms to process graph-structured data, the property of neighbouring datasets needs to be formally defined. The three main notions of DP on graphs can be formalised as follows:

\begin{definition}[\textit{Edge-level} DP]
Under \textit{edge-level} differential privacy, two graphs $\mathit{G_1}$ and $\mathit{G_2}$ are neighbouring if they differ in a single edge (either through addition or through removal of the edge) \cite{privateAnalysisGraph}. ($\varepsilon$-$\delta$) edge differential privacy is therefore preserved if equation \ref{eq:ed_dif_priv} holds for all events $S$ and all pairs of neighbours $\mathit{G}$, $\mathit{G'}$ that differ in a single edge. In this setting, two graphs $\mathit{G_1} = \{\mathit{V_1}, \mathit{E_1}\}$ and $\mathit{G_2} = \{\mathit{V_2}, \mathit{E_2}\}$ are neighbours if
\begin{equation}
    \mathit{V_2} = \mathit{V_1} \land
    \mathit{E_2} = \mathit{E_1} \setminus e_i, 
\end{equation}
where $e_i \in \mathit{E_1}$.
\end{definition}

\begin{definition}[\textit{Node-level} DP]
Under \textit{node-level} DP, two graphs $\mathit{G_1} = \{\mathit{V_1}, \mathit{E_1}\}$ and $\mathit{G_2} = \{\mathit{V_2}, \mathit{E_2}\}$ are defined as neighbouring if they differ in a single node and its corresponding edges (achieved through a node removal/addition) \cite{DPGraphsSlides}. ($\varepsilon$-$\delta$)-node differential privacy is therefore preserved if equation \ref{eq:ed_dif_priv} holds for all events $S$ and all pairs of neighbours $\mathit{G_1}$, $\mathit{G_2}$, that differ in a single node and its corresponding edges:
\begin{equation}
    \mathit{V_2} = \mathit{V_1} \setminus v_i \land
    \mathit{E_2} = \mathit{E_1} \setminus c,
\end{equation}
where $v_i$ is a node in $\mathit{V_1}$ and $c$ is the set of all edges connected to $v_i$.
\end{definition}

\begin{figure}[!ht]
    \centering
    \includegraphics[width=0.35\textwidth]{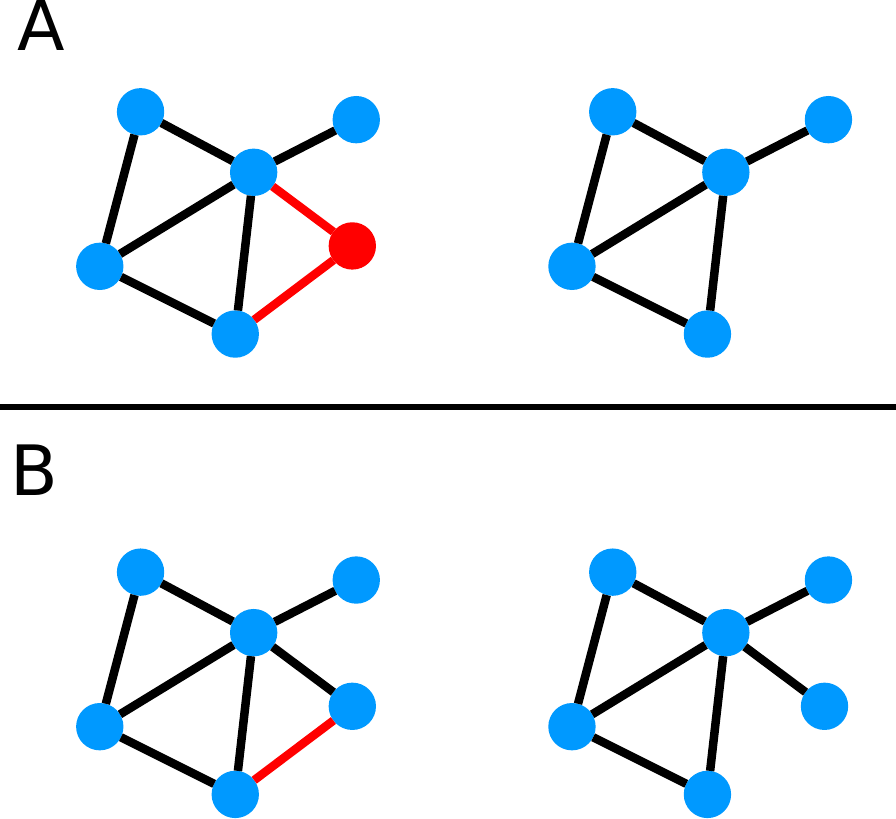}
    \caption{Two neighbouring graphs in the context of (\textbf{A}) \textit{node-level} DP and (\textbf{B}) \textit{edge-level} DP. By removing (\textbf{A}) one node and its adjacent edges or (\textbf{B}) one edge (displayed in red), two neighbouring graphs can be transformed into each other.}
    \label{fig:node_dp}
\end{figure}

Figure \ref{fig:node_dp} visualises these two main definitions of DP on graphs. Two neighbouring datasets (graphs) under \textit{node-level} DP and \textit{edge-level} DP are displayed in sub-figures \textbf{A} and \textbf{B}, respectively. 

For multi-graph datasets, we can define a different notion of privacy:

\begin{definition}[\textit{Graph-level} DP]
Under \textit{graph-level} DP, we define two multi-graph datasets $D_1 = \{G_{11}, G_{12}, \dots, G_{1n}\}$ and $D_2 = \{G_{21}, G_{22}, \dots, G_{2m}\}$ to be neighbours if they differ in one single graph (achieved through the addition or removal of one entire graph). ($\varepsilon$-$\delta$)-graph differential privacy is therefore preserved if equation \ref{eq:ed_dif_priv} holds for all events $S$ and all pairs of neighbouring datasets $D_1$ and $D_2$, where
\begin{equation}
    D_2 = D_1 \setminus G_{1i},
\end{equation}
and $G_{1i} \in D_1$.
\end{definition}

We now assume that $\mathcal{A}$ executes a function (or \textit{query}) $f$ over the graph dataset. When considering DP in GNNs, the function $f$ is a repeated composition of the forward pass, loss calculation, and gradient computation of the graph neural network. In order to determine the magnitude of noise that needs to be added, we are required to calculate the sensitivity of the function that noise is applied to. We will consider either the $L_1$- or the $L_2$-sensitivity of $f$.

\begin{definition}[$L_2$-sensitivity $\Delta$ of $f$]
Let $f$ be defined as above and $X$ be the set of all neighbouring databases. We can define the $L_2$-sensitivity of $f$ as:

\begin{equation}
    \Delta_2(f) := \max_{D, D' \in X, D \simeq D'} \Vert f(D)- f(D'))\Vert_2.
\end{equation}
When it is clear from context, we will omit the argument and write just $\Delta_{1/2}$. We note that the maximum is taken over all neighbouring pairs of datasets in $X$.
\end{definition}

Using the definition of $L_2$-sensitivity, we can formalise the Gaussian Mechanism on $f$:

\begin{definition}[Gaussian Mechanism]
Let $\Delta_2$ and $f$ be defined as above. The Gaussian Mechanism $\mathcal{M}$ is applied to the function $\mathbf{y}=f(x)$, $y \in \mathbb{R}^n$, as follows:
\begin{equation}
    \mathcal{M}(\mathbf{y}) = \mathbf{y} + \xi,
\end{equation}
\end{definition}

where $\xi \sim \mathcal{N}(0, \sigma \mathbb{I}^n)$. $\mathbb{I}^n$ is the identity matrix with $n$ diagonal elements and $\sigma$ is calibrated to $\Delta_2$.

Similarly to $L_2$-sensitivity, we can define the $L_1$-sensitivity as:

\begin{equation}
    \Delta_1(f) := \max_{D, D' \in X, D \simeq D'} ||f(D) - f(D')||_1.
\end{equation}

\begin{definition}[Laplace Mechanism]
Let $\Delta_1$ and $f$ be defined as above. The Laplace Mechanism $\mathcal{M}$ is applied to the output $\mathbf{y}=f(x)$, $\mathbf{y} \in \mathbb{R}^n$, as follows:
\begin{equation}
    \mathcal{M}(\mathbf{y}) = \mathbf{y} + \left (\xi_1, \xi_2, \dots, \xi_n \right),
\end{equation}
\end{definition}
where $\xi_i$ are I.I.D. draws from $\operatorname{Lap}\left( 0, \frac{\Delta_1}{\varepsilon} \right)$.




In general, one can furthermore distinguish between \textit{local} and \textit{central} DP. Under \textbf{local differential privacy (LDP)} \cite{xiong2020comprehensive} the data owner performs the noising step before the data reaches the analyst. Such interpretation can be preferable in low-trust collaborative learning settings, as no party other than its owner has access to the data before the learning task commences. Data owners only share a perturbed version of their training data, which reduces the amount of information an analyst can infer about the shared data itself, while still allowing to draw insights from the privatised aggregated data \cite{sajadmanesh2020locally}. Local DP thus bounds the information at the data source itself, minimising the potential privacy exposure \cite{kaissis2020secure}. An adversary is, therefore, unable to infer the input value with high confidence, but is possible to approximate the target query if provided with a large number of noisy samples \cite{sajadmanesh2020locally}. More detail about local DP on graph-structured data can be found in Section \ref{sec:ldp}.

When DP is, on the other hand, applied to the output of the computation instead of the input data, one speaks of \textbf{central differential privacy}. In this case, the noise is not added directly to the input data but instead to the computation outputs. Due to the properties of DP, only a bounded quantity of additional information can be derived about the data belonging to an individual, while the overall statistics of the whole dataset can still be approximately evaluated.

\section{Systematisation Methodology}
\label{sec:methodology}
We conducted a survey of papers that intersect the domains of graph analysis or deep learning on graphs with differential privacy. We employed the \textit{Google Scholar} and the \textit{Web of Science} search engines and examined papers that contained the keywords \mbox{\say{node-}}, \mbox{\say{edge-}}, \mbox{\say{graph-}} \mbox{\say{differential privacy}} between January, 2007 and February 2022. Our searches often had to be coupled (e.g. \say{node differential privacy graphs}), as notions such as \textit{graphs} or \textit{nodes} are often used in unrelated concepts such as computation graphs or network nodes. We selected 51 studies, which we partitioned based on the application of DP employed in each work: \textit{node-level} DP, \textit{edge-level} DP, \textit{graph-level} DP, and \textit{local DP} and separated the works into \textit{graph analysis} and \textit{GNN training} applications. We additionally recorded the contexts in which DP was applied. A summary of the works that we discuss in this study can be found in Table \ref{tab:summary}.

We observed that a large number of studies concentrate on the usage of graph datasets but explicitly not on the application of GNNs. The large amount of research in the context of DP on graphs in general shows the importance of applying differentially private algorithms to graph-structured data. However, applications of DP to GNNs is currently underrepresented, presumably due to the fact that GNNs are a relatively recent deep learning method, and the application of DP to GNNs entails several challenges. For example, there is no singular explicit notion of \say{DP} in different graph machine learning settings, as discussed below. Furthermore, the here-presented systematisation of different possibilities to apply differential privacy to graph neural networks will act as a comprehensive guide to practitioners and aid in the the development of new methods in this area. With the advent of privacy-preserving machine learning and the strong interest in geometric deep learning applications, we strongly believe the differentially private training of GNNs to be a promising future research area with several applications to sensitive data. Some exemplary application areas are discussed in Section \ref{sec:applications}.

\begin{table*}[!ht]
\centering
\resizebox{\textwidth}{!}{%
\begin{tabular}{@{}lccccclll@{}}
\toprule
 & \textbf{Edge-DP} & \textbf{Node-DP} & \textbf{Graph-DP} & \textbf{LDP} & \textbf{Year} & \textbf{Reference}  &\textbf{Context} & \textbf{$\varepsilon$} \\
 \midrule
  \multirow{45}{*}{\rotatebox[origin=c]{90}{\textbf{Graph Analysis}}} & \checkmark & \checkmark & & & 2007 & Nissim et al. \cite{nissim2007smooth}  & Estimation for spanning trees & -\\
  & \checkmark & & & & 2009 & Hay et al. \cite{hay2009accurate} & Graph degree estimation & $[0.01;1]$\\
  & \checkmark & & & & 2009 & Mir et al. \cite{mir2009differentially}  & Graph estimation  & -\\
  &  &  & & & 2011 & Gehrke et al.** \cite{gehrke2011towards} & Zero-knowledge statistics estimation & -\\
  & \checkmark & & & & 2011 & Machanavajjhala et al. \cite{machanavajjhala2011personalized}  & Privacy in social graphs & $[0.5;3]$\\
  & \checkmark & & & & 2011 & Sala et al. \cite{sala2011sharing} & Release of private graphs & $[0.1;100]$\\
  & \checkmark & & & & 2011 & Karwa et al. \cite{privateAnalysisGraph} & Private subgraph counting &$0.5$\\
  & \checkmark & & & & 2012 & Gupta et al. \cite{gupta2012iterative}  &  Synthetic data for graph cuts &-\\
  & \checkmark & & & & 2012 & Karwa et al. \cite{karwa2012differentially} &  Release of graph degree sequences &-\\
  & \checkmark & & & & 2012 & Mir et al. \cite{mir2012differentially} &  Private release of graph distribution &$0.2$\\
  & \checkmark & \checkmark &  & & 2013 & Blocki et al. \cite{blocki2013differentially} &  Restricted sensitivity for DP &-\\
  & & \checkmark & & & 2013 & Chen et al. \cite{chen2013recursive} &  Private graph database aggregation &$[0.1;0.5]$\\
  & & \checkmark & & & 2013 & Kasiviswanathan et al. \cite{kasiviswanathan2013analyzing} &  Private graph analysis &-\\
  & & & \checkmark & & 2013 & Shen et al. \cite{shen2013mining} &  Private graph pattern mining &$[0.1;1]$ \\
  & \checkmark & & & & 2013 & Wang et al. \cite{wang2013differential} & Private spectral graph analysis & $460$\\
  & \checkmark & & & & 2013 & Wang et al. \cite{wang2013preserving} & Private spectral graph analysis &-\\
  & \checkmark & & & & 2014 & Chen et al. \cite{chen2014correlated} & Correlated network data release & $[0.6;1]$\\
  & \checkmark & & & & 2014 & Lu et al. \cite{lu2014exponential} & Estimation of graph model parameters & $[0.1,1]$\\
  & & \checkmark & & & 2014 & Proserpio et al. \cite{proserpio2014calibrating} & Synthetic graph generation & $[0.01;10]$\\
  & \checkmark & \checkmark & \checkmark & & 2014 & Task et al. \cite{task2014should} &  Private social network analysis &-\\
  & & \checkmark & & & 2016 & Day et al. \cite{day2016publishing} &  Private graph distribution release &$[0.1;2]$\\
  & \checkmark & & & & 2016 & Jorgensen et al. \cite{jorgensen2016publishing} & Private attributed graph models &$[1;20]$\\
  & & \checkmark & & & 2016 & Raskhodnikova et al. \cite{raskhodnikova2016lipschitz} &  Private release of graph statistics & -\\
  & \checkmark & & & \checkmark & 2016 & Wang et al. \cite{wang2016using} &  Private aggregation of data & $[0;2]$\\
  & \checkmark & \checkmark & & \checkmark & 2017 & Qin et al. \cite{qin2017generating} & Private release of social graphs & $[0;7]$\\
  & \checkmark & \checkmark & & & 2017 & Zhu et al. \cite{zhu2017differential} &  Applications of differential privacy & - \\
  & \checkmark & \checkmark & & \checkmark & 2018 & Cormode et al. \cite{cormode2018privacy} & Private data release & - \\
  & & \checkmark & & & 2018 & Macwan et al. \cite{macwan2018node} & Private release of graph data & $0.5$\\
  & \checkmark & & & & 2019 & Arora et al. \cite{arora2019differentially}  &  Graph sparsification & -\\
  & & \checkmark & & & 2019 & Sealfon et al. \cite{ullman2019efficiently} & Estimation of graph statistics & -\\
  & & & & & 2019 & Sun et al. \cite{sun2019analyzing} & Subgraph statistics, decentralised DP & $[1;10]$ \\
  & & \checkmark & && 2019 & Yuxuan et al. \cite{yuxuan2019graph} &   Private histogram release &-\\
  & \checkmark & & & & 2020 & Chen et al. \cite{chen2020publishing} & Private synthetic data release & $[2;5]$\\
  & & \checkmark & & & 2020 & Liu et al. \cite{liu2020publishing} &  Node strength distribution  & $[0.1;2]$\\
  & & \checkmark & & & 2020 & Zhang et al. \cite{zhang2020community} &  Private social graph release & $[0.1;20]$\\
  & & \checkmark & & \checkmark & 2020 & Zhang et al. \cite{zhang2020differentially} &  Control-flow graph coverage analysis & $[2^{-5}; 2^5]$\\
  & & \checkmark & & & 2021 & Iftikhar et al. \cite{iftikhar2021dk} &  Private release of degree distribution & $[0.01;10]$\\
  & & \checkmark & & & 2021 & Fichtenberger et al. \cite{fichtenberger2021differentially} & Private dynamic graph algorithms & -\\
  & \checkmark & & & \checkmark & 2021 &  Imola et al. \cite{imola2021locally} &  Private sub-graph counting & $[0;2]$\\
  & & \checkmark & & & 2021 & Lan et al. \cite{lan2021sensitivity} & Private node strength histogram release & $[0.1;2]$ \\
  & & \checkmark & & & 2021 & Liu et al. \cite{liu2021graph} & Private degree histogram release  & $[0.1;2]$\\
  & & \checkmark & & & 2021 & Sealfon et al. \cite{sealfon2021efficiently} &  Private graph density estimation & -\\
  & \checkmark & \checkmark & & \checkmark & 2021 & Xia et al. \cite{xia2021dpgraph} & Benchmark platform for DP on graphs & - \\
  & \checkmark & & & \checkmark & 2021 & Zheng et al. \cite{zheng2021efficient}  & Private graph publication framework &-\\
  &\checkmark  & & & & 2021 & Zheng et al. \cite{zheng2021network} &  Network Generation & $[0.1;440]$\\
\midrule
\multirow{8}{*}{\rotatebox[origin=c]{90}{\textbf{GNNs}}} & \textbf{Edge-DP} & \textbf{Node-DP} & \textbf{Graph-DP} & \textbf{LDP} & \textbf{Year} &  \textbf{Reference} & \textbf{Context} & \textbf{$\varepsilon$}\\
\midrule
     & & & & \checkmark & 2020 & Sajadmanesh et al.* \cite{sajadmanesh2020locally} & Locally private GNNs & $[0.01;3]$\\
     & & \checkmark & & & 2021 & Daigavane et al. \cite{daigavane2021node} & Node-level DP in GNNs & $[5;30]$\\
     &  & & & & 2021 & Igamberdiev et al.* \cite{igamberdiev2021privacy} & Private text classification & $[1;100]$\\  
     & & & & & 2021 & Olatunji et al.* \cite{olatunji2021releasing} & Private GNN and graph data release & $[1;40]$\\  
     & & & & & 2021 & Zhang et al.* \cite{zhang2021graphmi} & Attacks on GNNs &$[1;10]$ \\  
     & & & \checkmark && 2022 & Mueller et al. \cite{mueller2022differentially} & Graph-level DP for graph classification & $[0.5;20]$\\
\end{tabular}
}
\caption{Summary of existing works on DP on graphs, ordered ascending by publication year and alphabetically within the same year. The works are split into \textit{Graph Analysis} and \textit{GNNs}. Ticks in columns \textbf{Edge-DP}, \textbf{Node-DP}, and \textbf{Graph-DP} specify which notion of privacy was used. A tick in column \textit{LDP} indicates that the authors used local DP. The asterisks (*) indicates that the DP notion is not clearly stated. Two asterisks (**) indicate the utilisation of zero-knowledge privacy. The column $\varepsilon$ reports the privacy budget that was evaluated in the respective works.}
\label{tab:summary}
\end{table*}  

\section{Exposition}
\label{sec:exposition}
In this section, we outline and discuss methods from the research field of differentially private graph analysis and graph machine learning. We identify and consider two separate lines of work: \textbf{(a)} DP in traditional graph analysis methods and \textbf{(b)} DP in graph neural networks. We therefore separate the works in Table \ref{tab:summary} depending on their association with one of those classes. We also indicate the notion of DP that was applied in the respective research in the columns \textit{Edge-DP}, \textit{Node-DP}, \textit{Graph-DP}, and \textit{LDP} and summarise ranges of the privacy budget $\varepsilon$ if they were reported in the respective works.
The line of work of DP in traditional graph analysis \textbf{(a)} includes methods for privately computing graph statistics like degree-distributions \cite{hay2009accurate}, frequent sub-graph-mining \cite{shen2013mining}, and sub-graph counting \cite{blocki2013differentially}, as well as private graph release \cite{mir2009differentially,jorgensen2016publishing,olatunji2021releasing}. 
The works of DP for GNN training \textbf{(b)} include, for instance, text classification \cite{igamberdiev2021privacy}, whole-graph classification \cite{mueller2022differentially}, and attacks on GNNs \cite{zhang2021graphmi}.


As indicated in Table \ref{tab:summary}, we generally observe a focus on \textit{edge-level} DP in earlier papers, compared to a more frequent utilisation of \textit{node-level} DP in more recent works. We attribute this to the fact that \textit{node-level} DP is more challenging to achieve, but offers stronger privacy guarantees (as it considers the privacy of a node and all its adjacent edges). Works on \textit{graph-level} DP are quite rare. However, we believe this notion of DP to be promising and given that different works name the same concept differently, we still included graph-level DP in Table \ref{tab:summary}.

We furthermore observe that, in the works discussing DP on GNNs, authors frequently omit to specifically assign the guarantees provided to one of the aforementioned DP notions, which highlights the need for more systematic approaches to defining DP in graph learning tasks. We attribute this lack of specification to missing systematisation of terminology in this area as well as the challenging task to differentiate the individual notions of privacy in graph learning tasks and their dependence on the dataset and the application area.

\subsection{DP on Graph-Structured Data} 
\label{sec:dp_graphs}
In this section we give an overview of existing DP notions on graphs, independent of context and task. 
The first application of differentially private computation on graph data was introduced by Nissim et al. \cite{nissim2007smooth}. Authors showed an estimation of the cost associated with the computation of a minimum spanning tree and triangle counts in a differentially private manner. In their work, the authors opted for the utilisation of \textit{edge-level} DP. 

As described above, ensuring data privacy on graphs presents additional challenges compared to structured databases such as image or tabular datasets, since the data points are inter-connected and the graph structure itself can contain sensitive information. Furthermore, depending on the application it can be desired to protect different parts of the graph.
One fundamental challenge is therefore the issue of sensitivity calculation. 

In cases of graphs, this value can be challenging to obtain as it depends not only on the structure of the graph but also on the attributes of the query function. Two main methods have been proposed to obtain node differentially private algorithms which are either based on (a) the utilisation of projections, for which sensitivity can be bounded, or (b) on computing Lipschitz extensions \cite{raskhodnikova2016lipschitz, blocki2013differentially, chen2013recursive}. Raskhodnikova et al. \cite{raskhodnikova2016lipschitz} study the efficient computation of Lipschitz extensions for multi-dimensional functions on graphs, which can be obtained in polynomial time, and determine that they do not always exist - in comparison to Lipschitz extensions for 1-dimensional functions. 

In the next sections, we give more details about the different definitions of DP on graphs in \textit{node-level}, \textit{edge-level}, \textit{graph-level} DP as well as some alterations and combinations of these, with respective interpretations of what is implied by neighbouring datasets in each setting.

\subsubsection{Edge-Level Differential Privacy} \label{sec:edge_privacy}
There exist several approaches that allow to release graph statistics with \textit{edge-level} DP guarantees, including sub-graph counts \cite{karwa2012differentially}, spanning tree estimation \cite{nissim2007smooth}, degree distributions \cite{task2014should, hay2009accurate} and graph cuts \cite{gupta2012iterative}. Those settings set a focus on privatising the relationships between nodes. This can be applied to social network graphs \cite{hay2009accurate, mir2009differentially} or location graphs \cite{xie2016learning}, where the edges contain sensitive information, but the data represented in the nodes of the graph are assumed to be publicly known or non-sensitive.

\subsubsection{Node-Level Differential Privacy} 
\textit{Node-level} differential privacy is a strictly stronger guarantee than edge-level differential privacy \cite{kasiviswanathan2013analyzing}. This is of particular importance in scenarios where graphs are very sparse, and thus, the removal of a single node can alter the graph structure severely. For instance, the number of triangles in a graph with $n$ nodes can increase by $\binom{n}{2}$ when inserting a single additional node. Consequently, these functions tend to have high sensitivity \cite{raskhodnikova2016lipschitz}, resulting in an unnecessarily large noise magnitude. Bounded-degree graphs (graphs where each node has an upper limit of edges and the degree of each node is therefore bounded) can assist in lowering the sensitivity. Here, the removal of a single node results in an upper-bounded change in edges which typically leads to a reduced impact on the output of the algorithm. When calculating the number of triangles in a graph, for instance, maximum change of a $D$-bounded-degree graph is $\binom{D}{2}$ which is strictly smaller than $\binom{n}{2}$ if $D < n$.

Settings that can benefit most from this formulations of DP are those that put an emphasis on the data within the node itself yet additionally privatise the connections between the nodes include studies on social networks \cite{qin2017generating, blocki2013differentially}, degree histogram distribution \cite{liu2021graph, iftikhar2021dk, macwan2018node}, and recommendation systems \cite{machanavajjhala2011personalized}. 

\subsubsection{Graph-Level Differential Privacy}
So far, \textit{graph-level} DP has not been explored in great detail, neither in the context of graph analysis nor in GNNs. Task et al. \cite{task2014should} name this notion of privacy \textit{partition privacy} and show its application to graph analysis of social networks. Shen et al. \cite{shen2013mining} investigate the mining of frequent graph patterns in multi-graph datasets and apply the mechanism of \textit{graph-level} DP to their algorithm. They use Markov Chain Monte Carlo (MCMC) random walks to discover frequently appearing sub-graphs in the graph dataset and infer graph statistics under graph-level DP.

In the context of GNN training, \textit{graph-level} DP can be applied in learning settings that investigate graph classification tasks, e.g. drug discovery or molecule classification \cite{duvenaud2015convolutional}, discovering disease-specific biomarkers of brain connectivity \cite{li2019graph,li2021braingnn}, or shape analysis \cite{wei2020view}. This way, privacy guarantees can be given to the individuals, whose sensitive information is contained in those multi-graph datasets. For instance, in the setting of drug discovery, a group of pharmaceutical companies can collaborate on a graph classification task, while bounding the information that can be inferred about their individual molecules, which represent the private data in this context. Mueller et al. \cite{mueller2022differentially} apply graph-level DP for classification tasks on several sensitive datasets, implementing the concept of graph-level DP on GNNs and showing potential applications.

\subsubsection{Further Definitions of DP on Graphs}
There exist additional notions of DP on graph-structured data that have not yet found a widespread adoption and are mostly derived from the notions formalised above. 

\paragraph{k-Edge Differential Privacy}
One such formulation is \textit{k-edge differential privacy} introduced by Hay et al. \cite{hay2009accurate}. It defines a stricter notion of \textit{edge-level} DP, where two graphs $\mathit{G}=\{\mathit{V}, \mathit{E}\}$ and $\mathit{G'}=\{\mathit{V'}, \mathit{E'}\}$ are neighbours if $\vert V \oplus V' \vert + \vert E \oplus E' \vert \leq k$. Hereby, $\oplus$ denotes the symmetric difference. If $k=1$, the definition recovers \textit{edge-level} DP. However, if $k = \vert V \vert$ \textit{k-edge-level} DP is a stricter definition than \textit{node-level} DP, as the set of neighbouring graphs in the definition of \textit{node-level} DP is a subset of the neighbouring graphs under \mbox{$k$-edge-level} DP. For nodes with a degree smaller then $k$, $k$-edge-level DP provides an equivalent protection as \textit{node-level} DP. Nodes with a degree $\geq k$ face more exposure, since they have more edges. However, one can argue that those high degree nodes have a higher impact on the general graph structure and it might therefore be necessary to expose them to larger privacy risks to allow analysts to accurately measure graph statistics. The authors experimentally evaluate their notion of k-edge-differential privacy on social network data from Flickr, LiveJournal, Orkut, and YouTube.

\paragraph{Out-Link Differential Privacy}
Another definition of DP on graphs was introduced by Task et al. \cite{task2014should} and is termed \textit{out-link differential privacy}. In this context directed graphs are considered, where it is possible to distinguish between incoming and outgoing edges of nodes. Under this notion, two datasets are considered to be neighbouring if all \textit{out-links} (outgoing edges) of an arbitrary node are added or removed. Formally, two graphs $\mathit{G_1} = \{\mathit{V_1}, \mathit{E_1}\}$ and $\mathit{G_2} = \{\mathit{V_2}, \mathit{E_2}\}$ are neighbours, if $\mathit{V_1} = \mathit{V_2}$ and $\mathit{E_2} = \mathit{E_1} - \{(v_1, v_2) | v_1=x\}$ for an $x \in \mathit{V_1}$. $(v_1, v_2)$ hereby defines an edge going from node $v_1$ to node $v_2$. 

\textit{Out-link} DP is strictly weaker then \textit{node-level} DP, but in many applications comparable to \textit{edge-level} DP. Under this notion of DP, an attacker would not be able to determine whether a person $x$ contributed their data to the construction of the graph and participants in the graph can hide their out-links. In the setting of a social network, for instance, a person $x$ can deny friendships. Others can still claim to be friends with person $x$, but the latter can deny that those connections are mutual (i.e. that person $x$ has out-going links to adjacent nodes). The authors argue that \textit{out-link} privacy simplifies sensitivity computation and reduces noise addition requirements, enabling queries that would be infeasible under previous DP definitions. 

Similar to $k$-edge-level DP, \textit{out-link} DP can also be extended to $k$-out-link privacy. In this case, neighbouring datasets are considered, that differ in $k$ out-links compared to the original dataset. When considering 2-out-link privacy, for example, two nodes can simultaneously deny all their out-links. This would also enable to protect a complete mutual edge, resulting in \textit{edge-level} DP in addition to \textit{out-link} DP.

\paragraph{Zero-Knowledge Privacy}
Gehrke et al. \cite{gehrke2011towards} introduce a stricter formulation of \textit{node-level} DP, namely \textit{zero-knowledge privacy} on graphs, which authors argue is particularly desirable in social network analysis. It relies on a notion similar to the one of cryptographic zero-knowledge proofs \cite{feige1988zero}, which entails that a protocol participant obtains a computation result with \say{zero additional knowledge} about the data used to perform this computation. A privacy mechanism $\mathcal{M}$ is (\textbf{Agg}, $\varepsilon$)-zero-knowledge private if there exists a simulator $\mathcal{S}$ and an $agg$ from the family of algorithms \textbf{Agg} such that for all neighbouring datasets $D_1$ and $D_2$ the following holds: $\mathcal{M}(D_1) \approx_{\varepsilon} \mathcal{S}(agg(D_2))$ \cite{gehrke2011towards}.

Authors of \cite{gehrke2011towards} apply this definition to ensure that a mechanism does not release additional information apart from \say{aggregate information} which is considered acceptable to release to ensure usability. 

\paragraph{Relationship Differential Privacy}
Imola et al. \cite{imola2021locally} introduce a notion called \textit{relationship DP}, a definition falling under local DP. Here, one edge in a graph is masked during the entire learning process. In a setting of social network analysis, relationship DP assumes that each user only knows their own connections (i.e. friends), requiring users to have a higher degree of \say{trust} when interacting with their immediate neighbours. Given two users $v_i$ and $v_j$ that share a link in the social network, under relationship-DP a user $v_i$ has to trust its adjacent user $v_j$ not to leak information about their shared connection. Intuitively, \textit{edge-level LDP} considers the edge from user $v_i$ to user $v_j$ and the edge from user $v_j$ to user $v_i$ to be two separate \say{secrets}, whereas relationship DP assumes that the two edges represent the same \say{secret}. (More details about edge-level LDP can be found in Section \ref{sec:ldp}.) Therefore, the trust model of relationship DP is a stronger one than the one of \textit{edge-level LDP}, which does not hold any assumptions about what other users do, but weaker than the one of centralised edge-level DP, where all edges are held by a centralised party. If a randomised algorithm $\mathcal{M}$ provides $\varepsilon$-edge-level LDP, then $\mathcal{M}$ provides $2\varepsilon$-relationship DP, given that an edge $(v_i, v_j)$ affects two elements in the adjacency matrix of the graph and the property of group privacy \cite{dwork2014algorithmic}.

The authors apply this formulation of privacy to algorithms for sub-graph, k-star, and triangle counting, which can be used to analyse connection patterns in graphs.

\paragraph{Edge-Weight Privacy}
For shortest path or distance queries on graphs, edge-level and node-level DP are not well suited, since both queries usually return a set of edges, which violates both edge-level and node-level DP. Therefore, Sealfon \cite{sealfon2016shortest} introduced a different notion of privacy on graphs: \textit{edge-weight privacy}. This notion of privacy is applicable if the edge weights of a graph contain private data, whereas the graph structure itself is publicly available and does not need to be protected. An example would be traffic data in a known street system.

\paragraph{Node Attribute Privacy}
Chen et al. \cite{chen2020publishing} define another notion of privacy for attributed graphs. An attributed graph $G = (V,E,X)$ is the set of vertices $V$, edges $E$ and node attributes $X$. In this definition of privacy, two graphs are defined to be neighbouring if they differ in one edge or in the attribute vector of one node. So in this scenario, the presence of nodes is assumed to be non-private, whereas the connections (edges) between the nodes as well as the attributes that define the nodes contain private information. This definition can for example be useful in social networks, where the existence of a profile can be publicly known but friendships and personal attributes (stored in the profiles/nodes) are private.

\subsubsection{Local DP on Graphs}
\label{sec:ldp}
There exist several works that target the preservation of local differential privacy (LDP) on graph-structured data. The advantage of local DP  \cite{kasiviswanathan2011can} in comparison to central DP is that no trusted third party is required. LDP can and has been applied to both, classical graph analysis and graph neural networks.
Qin et al. \cite{qin2017generating} define \textit{edge-level} and \textit{node-level} LDP in the context of neighbour lists. A neighbour list of a vertex $v_i$ in a directed graph with $n$ vertices is defined to be an n-dimensional bit vector $(b_1, \dots, b_n)$, where $b_i = 1$, $i \in [1; n]$, if and only if there exists an edge $(v_i, v_j)$, going from $v_i$ to $v_j$, in the graph, otherwise $b_i = 0$.
\textit{Edge-level LDP} is then defined for two neighbour lists that differ in exactly one bit, whereas \textit{node-level LDP} is defined for any two neighbour lists.

\paragraph*{Locally private graph analysis}  
Examples for LDP in graph analysis tasks include Zhang et al. \cite{zhang2020differentially}, who perform control-flow graph coverage analysis under \textit{node-level} LDP and Imola et al. \cite{imola2021locally}, who apply LDP to sub-graph counting, k-star and triangle counts while preserving \textit{edge-level} LDP.

\paragraph*{Locally private GNNs}
LDP can also be applied to GNNs, where settings such as decentralised social networks can benefit from this property, as shown by Sajadmanesh and Gatica-Perez \cite{sajadmanesh2020locally}. They introduce a privacy-preserving architecture-agnostic GNN algorithm, which preserves private node features under LDP. Their architecture includes an LDP encoder and an unbiased rectifier, which functions as the communicator between the server and the graph. This algorithm can be applied in a setting where either the node features or the labels (or in certain cases both) are to be kept private regardless of the GNN architecture. Authors use a so-called \textit{multi-bit mechanism} which allows the nodes to perturb their features before passing them to the server. The server then processes this noisy data through the first convolutional layer. GNNs aggregate the node features before passing them through the activation function, which can be used as a denoising mechanism to average out the noise that was injected into the node features in the first place. The authors employ a generalised randomised response mechanism \cite{kairouz2016discrete} to preserve privacy of node labels. However, they explicitly do not preserve \textit{node-level} or \textit{edge-level} DP but protect the privacy of node features and labels. This leaves the graph structure itself unprotected, which remains an open challenge in this context. 

\subsubsection{DP for Graph Neural Networks}
\label{sec:dp_gnns}
While the notion of DP on traditional graph analysis and statistics applications (particularly for private data release) is well established, there exist significantly fewer studies on differentially private GNN training. This can be attributed to multiple factors, one of them being the number of different GNN machine learning settings (e.g. single- and multi-graph settings). This renders the identification of a standardised method for differentially private GNN training significantly more challenging. Furthermore, GNN learning is not yet a fully established area of research, leaving a number of learning contexts unexplored. In this section we introduce two methods that have been used to achieve differentially private training on GNNs.

\paragraph{DP-SGD Training of GNNs}
One of the most common methods to perform differentially private training in (non-graph) machine learning is differentially private stochastic gradient descent (DP-SGD) \cite{abadi2016deep}. Here, a gradient descent step is privatised through bounding the gradient $L_2$-norm (clipping) and through the addition of calibrated noise, such that the output of the gradient calculation over two neighbouring datasets can --with high probability-- not be well distinguished. This concept is not limited to SGD and can be applied to other first-order optimisation techniques, e.g. Adam. 
In standard machine learning, the clipping in DP-SGD is applied to the backward pass of each individual data point to minimise the amount of noise that has to be added to the gradients. This method, naturally befitting structured databases with well-defined notions of what an \say{individual} gradient entails, does not seamlessly extend to graph machine learning in all cases. For graph classification tasks, for instance, each graph can be seen as an individual entity in a multi-graph dataset and, therefore, \textit{graph-level} DP can be seen as a natural formulation in these learning settings. Here the standard procedure of DP-SGD can be transferred from database queries to graph learning tasks, matching database entries (rows) with individual graphs. This has been shown in \cite{mueller2022differentially}. Even though graph-level DP has not been explored much in research so far, we believe this to be an interesting and promising research are with multiple application areas, for example in medical settings with population graphs or brain networks (see Sections \ref{sec:populationgraphs} and \ref{sec:brainnetworks}).

However, this approach is not directly transferable to GNNs in a single-graph setting, because the individual data points in a graph (where its nodes or edges) cannot be separated without breaking up the graph structure, which is essential to the message passing mechanism of GNNs. This not only precludes a notion of \say{per-sample} gradients, but also privacy amplification by sub-sampling, which states that a DP mechanism run on a random sub-sample of a population results in tighter privacy guarantees than when applied to the whole population \cite{balle2018privacy}. To counter this effect, Igamberdiev et al. \cite{igamberdiev2021privacy}, for instance, implement a graph splitting method, which partitions the graph into smaller batches to approximate sub-sampling amplification and apply DP mechanisms to graph neural networks. In general however, no universally valid method to assign one of the DP formulations discussed above to the application of DP-SGD in GNNs on single-graph datasets has been proposed. Sajadmanesh and Gatica-Perez \cite{sajadmanesh2020locally} address this problem by applying local DP on the node features, without protecting the graph structure and Olatunji et al. \cite{olatunji2021releasing} utilise teacher-student models to allow the differentially private release of GNNs.
However, we believe that the utilisation of DP-SGD in GNNs is still an open research question which needs to be explored in more detail.

\paragraph{Private Aggregation of Teacher Ensembles}
\label{sec:PATE}
Differentially private stochastic gradient descent is one of the most common methods to offer DP guarantees in machine learning. However, there are also alternative methods of privacy preservation in machine learning: private aggregation of teacher ensembles (PATE), introduced by Papernot et al. \cite{papernot2016semi}. It leverages an ensemble (a collection) of so-called teacher models that are trained on disjoint datasets containing sensitive data. These models are not published but instead used as teacher models for a separate student model. The student model cannot access any single teacher model nor the underlying data. It instead relies on a noisy voting algorithm performed across all teacher models to make a prediction. Olatunji et al. \cite{olatunji2021releasing} recently introduced a framework named \textit{PrivGNN}, which leverages PATE and incorporates it into a collaborative GNN training setting. The method requires two datasets: Labeled private data for the teacher model and unlabeled public data for the student model. The knowledge of the teacher model, trained on the private graph, is then transferred to the student model, trained only on the public graph in a differentially private manner. 

One notable limitation of this method is the reliance on a publicly available unlabeled dataset that can be utilised by the teacher model. In general, this is a rather strong assumption, particularly in contexts relying on scarce, private datasets, such as non-Euclidean medical data, hindering the widespread adoption of PATE as the means of differentially private training. In general, PATE could be considered a private student-teacher data labeling mechanism rather than necessarily representing a method for private collaborative training. This issue is compounded by a low utility of PATE in graph settings, as the physical separation of datasets in graph learning destroys structural information, significantly reducing the utility of the trained model \cite{olatunji2021releasing}.

\subsection{Application Areas for DP on Graphs}
\label{sec:applications}
In this section we discuss how our findings from above can be and have previously been applied to graph learning tasks in order to establish which formulations of DP are most suitable for each context, and give insights into a selection of potential application areas for DP on graph-structured data. Lastly, we provide an outlook on promising future research in those settings. We chose three distinct learning contexts to allow us to cover all commonly used formulations of DP on graphs (i.e. node-level, edge-level and graph-level DP). Overall, more contexts relying on sensitive (or proprietary) data can benefit from a formalisation of DP, such as drug discovery \cite{jiang2021could} or location-based learning \cite{kessel2010bigml}. We leave an in-depth investigation of privacy in these settings as future work.

\subsubsection{Social Networks}
One of the more well-researched areas of private learning on graphs concerns social graphs \cite{gehrke2011towards,sala2011sharing,blocki2013differentially,task2014should,qin2017generating}, where the personally-identifying information is contained in the nodes of the graph and/or in the edges, defining the interactions between individuals, that could potentially allow to uniquely identify them (e.g. when spatio-temporal data is published \cite{de2013unique}). As a result, there exist two sensible routes to perform private learning on such data: \textit{edge-level} DP to protect the connections to other individuals in the graph and prevent unique identification of users like in \cite{machanavajjhala2011personalized,privateAnalysisGraph,sala2011sharing} and \textit{node-level} DP to protect the data of each individual itself (as well as the outgoing edges) like in \cite{gehrke2011towards,mir2012differentially,blocki2013differentially,qin2017generating,liu2020publishing,zhang2020community}. Numerous works have previously been employed to allow private release of social graphs or their associated statistics \cite{mir2012differentially,task2014should,macwan2018node}. 

Sajadmanesh et al. \cite{sajadmanesh2020locally} utilise locally differentially private GNNs in the context of social networks. However, further applications of differentially private GNNs on social networks remain to be studied.

\subsubsection{Population Graphs}
\label{sec:populationgraphs}
A surge in the amount of patient information available to practitioners lead to an increased need for data structures that are able to systematise the information about each individual patient. Such approaches allow the practitioners to benefit from a structured representation of electronic health records, particularly from spatio-temporal patient data \cite{liu2015temporal}, leading to a widespread adoption of patient graphs \cite{barbiero2021graph,duplaga2004universal,muller1996graph}. These data structures allow to encapsulate the information about patients across multiple departments and time periods, leveraging much more relevant information and leading to better predictions. One such scenario could involve representing each patient as a node and the whole patient population/cohort by a graph comprising the individuals, as described e.g. in \cite{gao2020mgnn,mao2019medgcn}. Connections between patients can, for instance, be based on their similarity (like in \cite{parisot2018disease}). An advantage of creating such patient population graphs for the application of DP mechanisms is that the graph can be explicitly degree-bounded, limiting the impact of individual nodes on the graph structure.

Alternatively, each node can be patient-specific data about a single individual collected at different times by various specialists. Either of these contexts, as they are relying on extremely sensitive data contained in each node, would benefit from the utilisation of \textit{node-level} DP in order to quantify and limit the amount of information revealed when node-level data is processed or released. 

\subsubsection{Brain Networks} 
\label{sec:brainnetworks}
Multi-graph settings present additional challenges both technically and conceptually, making formulation of DP in this learning context a difficult task. In such setting it is not the information contained in a single node or even in an inter-node connection that needs to be kept private, but rather the information contained in a graph as a whole. One prime example of such dataset that contains sensitive information on a whole-graph level, rather than on the level of its individual constituents is a brain network graph \cite{bullmore2011brain}. Such data is used extensively in neuroimaging problems \cite{yu2015assessing, sporns2012simple, menoret2017evaluating}. However, similarly to most medical datasets, due to the difficulty of obtaining such data (both because of the complexity of the task as well as of the privacy concerns) it is essential that the learning task is augmented with a suitable privacy-preservation mechanisms. In the case of brain network graphs, information about the value of individual voxels, or single connections to other voxels in the brain network are not necessarily personally identifying. Nonetheless, a collection of such interconnected points is considered to be a particularly sensitive medical dataset and it thus needs to be protected. For this setting, \textit{graph-level} DP is a particularly suitable technique for data release. To date, there only exists a small number of such implementations of differentially private multi-graph learning and we envision that such formulation can gain significance as part of the future work in the area. We recall that DP deep learning on brain graphs (with learning tasks similar to \cite{richiardi2013machine} for instance) can be implemented through a straightforward utilisation of DP-SGD, similarly to Euclidean contexts.

\section{Challenges and Outlook}
\label{sec:outlook}
In this section, we discuss a number of challenges associated with differentially private graph analysis, some of which can be attributed to the inter-connected nature of graphs, while others are inherent to DP itself. Note that we also discuss a number of potential complications arising in DP GNN training as well as in DP graph analysis and data release.

\subsection{Privacy Accounting}
Typically, in differentially private machine learning settings privacy loss can be bounded per individual data point (i.e. per image or table record), thus considering data points independently from each other, simplifying privacy loss accounting. However, due to an intrinsic inter-dependency of nodes in a graph, independence cannot be guaranteed and therefore quantifying the contribution of each individual becomes non-trivial.

Thus, there arises a need for concrete definitions which would allow the data owner(s) to determine the exact formulation of differentially private training that is applicable in the specific application areas. As noted by \cite{DPGraphsSlides}, the guarantees given by \textit{edge-level} DP and \textit{node-level} DP have different implications, which are based on the exact features data owners wish to protect. 

DP is inherently compositional, that is, DP algorithms composed with each-other yield a DP algorithm \cite{dwork2014algorithmic}. However, the heterogeneous composition of different formulations of DP in a graph setting has not been studied previously. We foresee that research in graph analysis and graph learning tasks will benefit from a more systematised approach to differentially private methods, like the ones we attempt to provide in this work. 

As we discussed above, attacks on GNNs typically result in greater attack success for the adversary, as graphs contain more information revealing the individuals and their relationships. Consequently, we see an open area of research that would bridge these two challenges of interpretation of suitable DP formulations and attack mitigations in the context of inference and inversion attacks. One such approach can be employed in large sensitive datasets, where it is not always desirable to run differentially private learning on the entire dataset at once. Reliance on random sampling of population from the dataset can result in not only a better performance, but additionally much tighter privacy bounds, providing higher privacy guarantees for the participants \cite{balle2018privacy}.

An orthogonal approach is followed by techniques aiming to account for \textit{individual privacy loss} \cite{feldman2021individual}. Here, a \say{bespoke} privacy guarantee is given to each individual participating in a computation, typically combined with a method to automatically terminate their participation when their individual privacy budget is exhausted. As the process of deciding to continue or halt a computation by considering the currently spent privacy budget is an instance of \textit{fully adaptive} composition, additional mechanisms are introduced: a privacy odometer (which tracks the privacy expenditure in the process of computation, without having to specify a privacy budget in advance) and the privacy filter (which stops the computation once the privacy budget is exceeded). The combination of these tools allows for a finer-grained control of the information that can be learned from each individual data point and potentially higher utility. The ability to compute individual privacy loss can allow a selective removal of individual nodes (and their corresponding edges), resulting in a much finer control of individual privacy expenditure. This method can permit tighter privacy bounding in settings where amplification by sub-sampling is not possible. However, it is also limited in applicability whenever the notion of a single individual within the graph is ill-defined.

\subsection{Privacy-Utility Trade-Off}
As we briefly discussed in section \ref{sec:exposition}, DP in general adversely affects the utility of the model or of the results derived from a differentially private graph analysis. Utility if often measured by the accuracy of a query or with similar evaluation metrics. Therefore, similar to differentially private machine learning on Euclidean data or release of statistics derived from the sensitive data, there persists an issue of \textit{privacy-utility trade-off}. This implies that the more \say{private} the result of the computation is (e.g. the lower the value of epsilon is), the less useful information can be inferred from that result not just by the adversary, but also by the end user of the trained model, potentially hindering the scientific progress based on the insights that could have otherwise been obtained from the study. This is further exacerbated by the inter-connected nature of the graphs, as it is not possible to guarantee independence of individual nodes, as we discussed above. Therefore, operations that limit the amount of information that can be derived from these nodes (e.g. through DP statistics release) affect not just the individuals, but also additional nodes connected to them. Thus, the utility loss can become more problematic when compared to datasets with independent data points and inflict additional penalties on the results of the computation. We note that this discussion is relevant to both graph datasets and GNNs, as the nature of GNN learning can only make full use of the data if these properties of graphs are preserved. Relying on GNN models pre-trained on publicly available data (similar to \cite{hu2020gpt,hao2021pre,qiu2020gcc}) could severely reduce the negative impact that DP has on utility, when used in transfer learning contexts. Here a model is trained on public data and subsequently fine-tuned on private data where higher privacy can be achieved, while having better utility. This approach was demonstrated in \cite{abadi2016deep} and more recently in \cite{tramer2021differentially} for non-graph machine learning tasks, demonstrating that --whereas training to the same utility \textit{from scratch}-- requires about one order of magnitude more data, results comparable to non-private training can easily be achieved by transfer learning.

\subsection{Computational Performance}
Beyond the aforementioned trade-offs in model generalisation performance, the utilisation of differential privacy is also associated with a computational performance overhead when employed in deep learning settings. This can be attributed to a requirement for per-sample gradient calculation, imposing a significant burden on model performance at train time. Moreover, due to noise addition and gradient clipping, models typically converge more slowly, thus prolonging the required training time \cite{kurakin2022training}.

\subsection{Interpretability of DP in Graphs}
DP can often be difficult to reason over from the perspectives of fairness \cite{farrand2020neither} and explainability \cite{dwork2019differential}. Moreover, its correct application is complicated by the introduction of unintuitive parameters like $\varepsilon$ or $\delta$ \cite{cummings2021need, kaissis2021unified}, or by the requirement to understand additional DP definitions like node, edge or graph-level DP. Thus, besides systems which automate sensitivity calculations and the application of DP to generic machine learning workflows \cite{usynin2021automatic}, works similar to \cite{cummings2019compatibility} are required, which investigate user expectations and interpretations of DP, paving the way for an improved \textit{user experience} for practitioners. 

Interpretability of GNNs in general is a highly discussed task in literature. The authors in \cite{mueller2022differentially} use an explainability method called \textit{GNNExplainer} \cite{ying2019gnn} to visualise and quantify the similarity between graph neural networks trained with and without DP-SGD to evaluate whether the privately trained network considers the same edges in the graph as important as the network trained with standard ML. We see potential in methods like these to get a better insight into differenitially private GNNs and increase their interpretability.

\subsection{Synthetic Graph Generation}
One final graph learning context that still remains an open challenge is private synthetic graph generation. The ability to generate synthetic samples allows one to augment existing datasets with additional data points in a privacy-neutral way, resulting in more diverse data representations. This, in turn, improves utility of the model trained on this data as well as empirically reduces the effectiveness of inference attacks \cite{paul2021defending}. There exist prior works in the area \cite{zheng2021network,wang2013differential,mir2009differentially,qin2017generating,karwa2012differentially} that allow to generate graph-structured data in a private manner, however, authors outline a number of limitations. Firstly, the effect of privacy-utility trade-off is much more profound in graph generation tasks, forcing the model owner to either lower the privacy guarantees or to generate graphs of much lower utility. Secondly, the number of DP formulations that are applicable to synthetic graph generation is rather limited: To-date, stronger privacy formulations like node-level DP are not yet widespread in the setting of private synthetic graph generation. Chen et al. \cite{chen2020publishing}, e.g., explore synthetic garph generation of social graphs under \textit{edge-level} DP. Gupta et al. \cite{gupta2012iterative} introduce a method for synthetic graph generation specifically tailored to graph cuts. Additionally, graph generation has so far been limited to simple benchmark datasets and has not been widely investigated under the lens of the privacy-utility trade-off in more challenging contexts. Qin et al. \cite{qin2017generating}, e.g., therefore resort to LDP to generate synthetic decentralized social graphs. Since this particular application of private graph-based learning is relatively new, we identify this to be a promising area of future work in the graph domain.

\section{Conclusion}
\label{sec:conclusion}
In this work, we explore and systematise applications of differential privacy in graph analysis and on graph neural networks. We discovered 51 works that perform differentially private data processing of graph structures, which we classify by the DP formulations employed in each work and summarise our findings in Table \ref{tab:summary}. We identify three main DP formulations with regards to the attributes of graphs considered to be sensitive: (1) edge-level, (2) node-level, and (3) graph-level differential privacy. We additionally discuss machine learning tasks (in particular those relying on GNNs) that require utilisation of sensitive graph-structured data and could hence benefit from a formalisation of differentially private learning. Subsequently, we discuss the limitations of DP when applied to such learning contexts, some of which are inherent to the choice of DP learning setting and some attributable to the inter-connected nature of graph structures specifically. We conclude our discussion with an analysis of graph learning tasks on sensitive data, summarise which DP formulations are suitable for different learning problems and identify promising areas of future research. We hope that our work offers practitioners a helpful overview of the current state of DP employed in graph-based learning, and will stimulate both foundational and application-focused future research.

\bibliographystyle{unsrt}  
\bibliography{literature}  



\end{document}